%% file: main.tex
\documentclass[sigconf]{acmart}

\usepackage{amsmath}
\usepackage{multirow}
\usepackage{graphicx}

\usepackage{colortbl}
\usepackage[table]{xcolor}
\usepackage{pgf}

\newcommand{\databg}[1]{%
  \begingroup
    \pgfmathsetmacro{\value}{#1}
    \pgfmathsetmacro{\absvalue}{abs(\value)}
    \pgfmathtruncatemacro{\percent}{min(75, round(75*\absvalue))}
    \pgfmathtruncatemacro{\sign}{ifthenelse(\value>0,1,ifthenelse(\value<0,-1,0))}
    \ifnum\sign=1
      \edef\mycolor{blue!\percent!white}%
      {\expandafter\cellcolor\expandafter{\mycolor}#1}%
    \else
      \ifnum\sign=-1
        \edef\mycolor{red!\percent!white}%
        {\expandafter\cellcolor\expandafter{\mycolor}#1}%
      \else
        {\cellcolor{white}#1}%
      \fi
    \fi
  \endgroup
}

\renewcommand\footnotetextcopyrightpermission[1]{} % removes footnote with conference information in first column

\begin{document}

    \title{Open the Oyster: Empirical Evaluation and Improvement of Code Reasoning Confidence in LLMs}
    
    \author{Shufan Wang}
    \affiliation{%
      \institution{Chu Kochen Honors College,\\Zhejiang University}
      \city{Hangzhou}
      \country{China}}
    \email{shufanwang@zju.edu.cn}

    \author{Xing Hu}
    \affiliation{%
      \institution{The State Key Laboratory of Blockchain and Data Security,\\Zhejiang University}
      \city{Hangzhou}
      \country{China}}
    \email{xinghu@zju.edu.cn}

    \author{Junkai Chen}
    \affiliation{%
      \institution{Singapore Management University}
      \country{Singapore}}
    \email{junkaichen@smu.edu.sg}
    
    \author{Zhiyuan Pan}
    \affiliation{%
      \institution{The State Key Laboratory of Blockchain and Data Security,\\Zhejiang University}
      \city{Hangzhou}
      \country{China}}
    \email{zy_pan@zju.edu.cn}

    \author{Xin Xia}
    \affiliation{%
      \institution{College of Computer Science and Technology,\\Zhejiang University}
      \city{Hangzhou}
      \country{China}}
    \email{xin.xia@acm.org}
    
    \date{\today}

    \input{sections//0abstract.tex}


    \maketitle

    \input{sections//1introduction.tex}

    \input{sections//2background.tex}

    \input{sections//3methodology.tex}

    \input{sections//4setup.tex}

    \input{sections//5results.tex}

    \input{sections//6discussion.tex} 

    \input{sections//7threats.tex}

    \input{sections//8related.tex}

    \input{sections//9conclusion.tex}

    \newpage

    \bibliographystyle{ACM-Reference-Format}
    \bibliography{references.bib}

\end{document}

%% file: sections/0abstract.tex
\begin{abstract}

    With the widespread application of large language models (LLMs) in the field of code intelligence, increasing attention has been paid to the reliability and controllability of their outputs in code reasoning tasks. Confidence estimation serves as an effective and convenient approach for evaluating these aspects.
    This paper proposes a confidence analysis and enhancement framework for LLMs tailored to code reasoning tasks. We conduct a comprehensive empirical study on the \textbf{confidence reliability} of mainstream LLMs across different tasks, and further evaluate the effectiveness of techniques such as prompt strategy optimisation and mathematical calibration (e.g., \textit{Platt Scaling}) in improving confidence reliability. Our results show that \textit{DeepSeek-Reasoner} achieves the best performance across various tasks, outperforming other models by up to $0.680$, $0.636$, and $13.652$ in terms of ECE, Brier Score, and Performance Score, respectively. The hybrid strategy combining the reassess prompt strategy and Platt Scaling achieves improvements of up to $0.541$, $0.628$, and $15.084$ over the original performance in the aforementioned three metrics. These results indicate that models with reasoning capabilities demonstrate superior confidence reliability, and that the hybrid strategy is the most effective in enhancing the confidence reliability of various models.
    Meanwhile, we elucidate the impact of different task complexities, model scales, and strategies on confidence performance, and highlight that the confidence of current LLMs in complex reasoning tasks still has considerable room for improvement. This study not only provides a research foundation and technical reference for the application of confidence in LLM-assisted software engineering, but also points the way for future optimisation and engineering deployment of confidence mechanisms.
    
\end{abstract}

%% file: sections/1introduction.tex
\section{Introduction}
    Recently, Large Language Models (LLMs) have shown outstanding abilities in code understanding, generation, and reasoning, which greatly enhances development efficiency and code quality.
    Developers have increasingly relied on LLMs to solve code reasoning tasks to assist with code review, debugging, refactoring, and testing~\cite{lo2023trustworthysynergisticartificialintelligence}. This trend not only drives a paradigm shift in software engineering but also raises higher demands for the interpretability and reliability of model outputs.

    Code reasoning tasks require the model to analyse a given code snippet at the syntactic, semantic, and logical levels. The model needs to infer aspects such as the input and output, runtime behaviour, branch paths, or variable values. As a core capability in automated code analysis, code reasoning underpins many practical applications in software engineering.
    In real-world scenarios, however, it is not enough for models to simply provide answers; developers also need to understand how much trust they can place in these outputs. Providing confidence information alongside model predictions allows developers to better assess the reliability of the results and make more informed decisions during software development and maintenance.
    
    \textbf{Confidence} refers to the model’s subjective probabilistic assessment of the correctness of its own output. 
    For example, when an LLM answers \textit{``the output of this code is 42''} with 90\% confidence, it means that the model believes there is a 90\% chance its answer is correct. 
    Confidence not only reflects the model’s self-awareness, but also provides developers with an important reference for judging the reliability of model outputs.  On one hand, current LLMs may still exhibit confident but incorrect behaviour in code reasoning, directly affecting the reliability of developers’ decisions. On the other hand, the reliability of confidence is directly related to the usability and trustworthiness of LLMs in practical software engineering. Therefore, studying the confidence of LLMs in code reasoning tasks is of great significance. Systematically analysing and improving the reliability of LLM confidence is of practical value for promoting the real-world application of intelligent code assistants.

    However, despite significant progress in code reasoning tasks, the reliability of LLM outputs still faces many challenges. 
    First, as probabilistic generative models, LLMs often produce uncertain outputs when they lack explicit knowledge or encounter complex logic. In some cases, they may even exhibit confidently incorrect predictions~\cite{jesse2023largelanguagemodelssimple}. Second, in real-world applications, developers urgently require a mechanism to quantify the trustworthiness of model outputs, so as to effectively manage high-risk or critical tasks and improve the safety and efficiency of human–AI collaboration. Confidence, as a core metric for measuring a model’s self-awareness and output reliability, should be an important dimension for evaluating LLMs’ code reasoning capabilities. 
    Nevertheless, current mainstream LLMs still exhibit significant shortcomings in confidence generation, calibration, and interpretability, making it difficult to provide developers with high-quality decision support.

    Recent studies have been proposed to analyse confidence for LLMs in natural language understanding~\cite{xiong2024can}, factual question answering~\cite{jiang2021knowlanguagemodelsknow}, arithmetic reasoning~\cite{kadavath2022languagemodelsmostlyknow}, and completion~\cite{spiess2024calibrationcorrectnesslanguagemodels}. 
    Although these studies offer valuable insights into the confidence mechanisms of LLMs, systematic reliability analysis of confidence in code reasoning remains relatively scarce. 
    Most existing work relies on objective metrics such as accuracy or F1 score, neglecting the quantification and optimisation of model self-awareness.
    This fails to meet the demand for highly reliable and interpretable models in engineering practice.
    Moreover, empirical studies on effective technical approaches to improve LLM confidence reliability, especially in code reasoning tasks, are extremely limited.

    To fill this research gap, this paper focuses on the confidence reliability of LLMs in code reasoning tasks. 
    We first systematically evaluate the confidence performance of mainstream LLMs on code reasoning benchmarks through multi-dimensional and fine-grained empirical analysis. Then, we explore different techniques, including \textbf{prompt strategy optimisation} and \textbf{mathematical calibration} (such as \textit{Platt Scaling}) to enhance the confidence reliability of LLMs in code reasoning. %Specifically, we not only conduct comparative analyses across different models, tasks, and confidence generation methods, but also systematically assess the effectiveness of \textbf{Prompt-Based Post-Processing} and \textbf{mathematical calibration} (such as \textit{Platt Scaling}) in improving confidence reliability.   
    This study reveals the capability boundaries, influencing factors, and improvement pathways for LLM confidence reliability, providing a solid data foundation and insightful mathematical analysis practice for subsequent research and practical applications.

    Specifically, we employed mathematical metrics such as \textbf{confidence calibration error} to evaluate the reliability of confidence exhibited by mainstream large language models on the benchmarks of relevant code reasoning tasks. Experimental results indicate that there are significant differences in the confidence reliability of different LLMs on code reasoning tasks. For example, the confidence calibration error of \textit{DeepSeek-Reasoner} on input-output prediction tasks is reduced by more than 50\% compared to the non-reasoning model, i.e., \textit{DeepSeek-Chat}.
    It demonstrates that reasoning ability markedly enhances confidence and reliability. 
    However, after mathematical calibration (such as \textit{Platt Scaling}), 
    %some models exhibit several-fold improvements in confidence intelligence. For instance, 
    \textit{DeepSeek-Chat}’s intelligence metric turns from negative to positive on certain tasks, with an improvement exceeding 100\%. Additionally, 
    %models that previously performed poorly, such as 
    \textit{Qwen3-1.7B} and \textit{GPT-3.5-Turbo} show a 20\% to 40\% reduction in confidence error after calibration, which significantly alleviates systematic bias.
    Overall, models with strong reasoning capabilities and large-scale architectures have a clear advantage in confidence reliability over traditional models. With appropriate calibration and optimisation, the confidence reliability of mainstream LLMs can be substantially improved. However, in complex reasoning tasks or low-performance model scenarios, confidence performance still falls short of practical engineering requirements. This suggests that while current LLM confidence mechanisms have demonstrated promising potential for improvement, further optimisation is needed to meet the high reliability demands of real-world development.

    The main contributions of this paper are as follows: 
    \begin{itemize}
        \item We propose and implement a systematic analytical framework for the confidence reliability of LLMs in the context of code reasoning tasks. On this basis, we conduct a comprehensive and quantitative \textbf{empirical study} on the confidence reliability of mainstream LLMs.
        \item We systematically evaluate the effectiveness of \textbf{prompt strategy optimisation} and \textbf{mathematical calibration} methods, revealing their applicability and limitations across different models and tasks.
        \item We provide an in-depth analysis of the impact of confidence reliability on practical software engineering applications, and offer feasible suggestions for the optimisation and engineering deployment of LLM confidence mechanisms.
    \end{itemize}

    \noindent \textbf{Paper Organization}. Section~2 introduces the background of this study. Section~3 provides a detailed account of the methodology adopted in this research. Section~4 presents the experimental design. Section~5 elaborates on the research results in detail. Section~6 outlines the implications of this study, with a discussion focus on the balance between optimising performance and practical utility. Section~7 highlights potential threats to the validity of the research results. Section~8 reviews related work. Finally, Section~9 summarises the conclusions of this study and highlights its implications for future research directions.

    \begin{figure*}
        \centering
        \includegraphics[width=\textwidth]{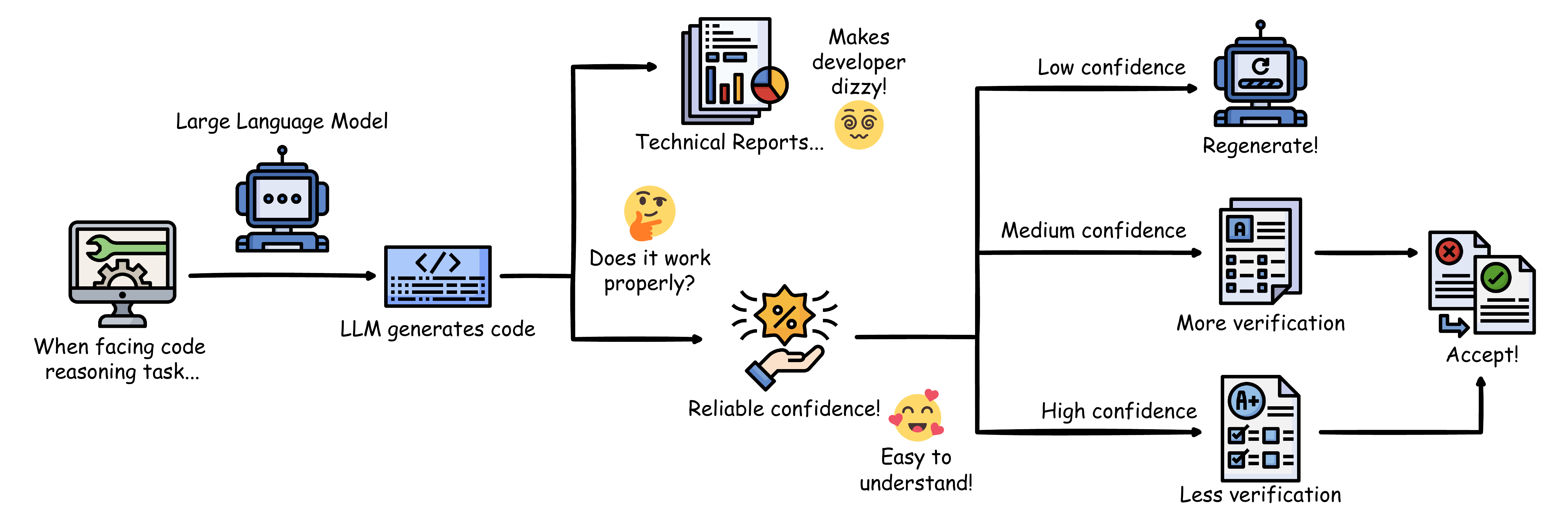}
        \vspace{-0.2cm}
        \caption{A flowchart illustrating the superiority of confidence, compared to traditional technical reports, in evaluating LLM-generated code for code reasoning tasks.}
        \vspace{-0.3cm}
        \label{fig:advantage}
    \end{figure*}

%% file: sections/2background.tex
\section{Background}

    This section discusses the background information relevant to this study, including \textbf{code reasoning} and \textbf{confidence}. The motivation for this research is also introduced in this section.

    \subsection{Code Reasoning}

        \textbf{Code reasoning} tasks aim to predict the behaviour of a programme without executing it by using LLMs~\cite{chen2025reasoning}. In the field of software engineering, code reasoning tasks possess extensive practical value. Tasks such as code review, debugging, and fault localisation all centre around code reasoning.

        Recent studies have emerged applying LLMs to code reasoning tasks. Min \textit{et al.}~\cite{min2024beyond} investigated the self-consistency of LLMs across different code reasoning tasks. 
        Miceli-Barone \textit{et al.}~\cite{micelibarone2023largerareharderfail} examined the extent of LLMs’ understanding of programming. Wu \textit{et al.}~\cite{wu2024reasoningrecitingexploringcapabilities} explored the transferability of LLMs’ code capabilities across different tasks. These studies mainly focus on the operational mechanisms of LLMs in code reasoning tasks and evaluate their performance.

       In addition, a variety of benchmarks for code reasoning tasks with LLMs have also been introduced. \textit{CRUXEval}~\cite{gu2024cruxevalbenchmarkcodereasoning} comprises synthetically generated simple Python programmes and their corresponding input/output pairs, with an emphasis on evaluating the ability of LLMs to predict inputs and outputs. 
       The improved benchmark, \textit{CRUXEVAL-X}~\cite{xu2025cruxevalxbenchmarkmultilingualcode}, extends the tasks to a wider range of programming languages and test samples. 
       Chen et al.~\cite{chen2025reasoning} propose \textit{REval}, which is designed to assess the ability of LLMs to predict dynamic execution properties, such as output prediction, branch prediction, and intermediate variable value prediction. \textit{CodeMind}~\cite{liu2025codemindevaluatinglargelanguage} aims to measure the code reasoning abilities of LLMs through both explicit and implicit tasks, such as independent execution reasoning, specification reasoning, and dynamic semantics reasoning. 
       In summary, these benchmarks provide comprehensive and multifaceted evaluation criteria for LLMs in code reasoning tasks.

    % xxx
    \subsection{Confidence}
        Confidence, also referred to as the probability of correctness as self-assessed by LLMs, denotes the subjective estimation by LLMs regarding the correctness of their own outputs when solving specific tasks. This metric originates from human cognitive science.~\cite{cosmidesS1996arehumansgoodintuitive}
        Just as humans tend to have a prior expectation about their ability to solve a task correctly, this concept can also be applied to the study of LLMs. Doing so helps to reveal the intrinsic reasoning mechanisms of LLMs. It may also contribute to enhancing their task-solving performance in the future.

        Some empirical studies on the metric of confidence have been proposed recently. Xiong \textit{et al.}~\cite{xiong2024can} and Jiang \textit{et al.}~\cite{jiang2021knowlanguagemodelsknow} focus on the confidence of LLMs in factual questions and arithmetic reasoning tasks. 
        %Subsequently, this research approach was extended to code-related tasks. 
        Spiess \textit{et al.}~\cite{spiess2024calibrationcorrectnesslanguagemodels} analyse confidence in code synthesis, code completion, and programme repair tasks. These studies not only examine the accuracy of LLMs' confidence in specific tasks but also provide some heuristic attempts at confidence calibration.
         With the emergence and widespread adoption of LLM-based code assistants such as GitHub Copilot, Cursor, and Tabnine in practical development~\cite{ziegler2022productivityassessmentneuralcode}, it is urgent to evaluate the correctness of the LLM-generated code~\cite{jesse2023largelanguagemodelssimple, lo2023trustworthysynergisticartificialintelligence}. These works also point out the potential of the concept of confidence in real-time evaluation during actual software development.

    \subsection{Motivation}

        As mentioned previously, humans tend to have a prior expectation regarding their ability to solve a given task, and accordingly adjust their problem-solving strategies.
        In other words, the subjective factor of \textbf{confidence} can effectively and positively alter our objective strategies, thereby enabling us to achieve the most efficient path throughout the entire task-solving process. This is equally applicable to LLMs in software development.

        Figure~\ref{fig:advantage} illustrates a scenario where developers reason about the programme’s execution process for complex programmes by using LLMs. %they need to reason about the programme’s execution process.   
        %Due to the complexity of the programme, they turn to an LLM-driven code agent for assistance. 
        The LLM can quickly generate a reasoning result, but the developer remains sceptical about the correctness of the LLM’s reasoning. 
        One approach is to evaluate the generated code using traditional code analysis tools. However, this requires developers to read complex technical reports for assessment, and the complexity of this process negates the convenience of using LLMs to assist with tasks. Alternatively, if the developer is able to obtain a \textbf{reliable} confidence score from the LLM regarding its own output through appropriate methods, then they can take appropriate action in response. When confronted with low confidence, the current result is simply discarded and the LLM is prompted to regenerate. In cases of moderate or high confidence, the former requires more extensive verification before being accepted, whereas the latter can be accepted with less verification.

        The advantages of this approach is mainly reflected in the following two aspects. 
        First, consider the method for obtaining this confidence score. If it falls within the approaches explored in this study, which include prompting the LLM to generate a confidence score under specific prompts together with simple mathematical processing, then the process is undoubtedly convenient. It also incurs less additional cost compared to the code reasoning task itself.
        Second, compared to complex technical reports or verbose feedback, the concept of confidence is inherently more intuitive and easier for humans to understand, thereby saving developers considerable time in assessing the reliability of task results. %Through such an example, we have reason to believe that obtaining a reliable \textbf{confidence} score alongside code reasoning outputs from LLMs can indeed benefit practical software engineering development.

       %Figure~\ref{fig:advantage} vividly illustrates this process. This approach is conducive to the efficient application of LLMs in the field of software engineering.

        However, this raises the question: how can we determine whether the confidence generated by an LLM is \textbf{reliable}? 
        Intuitively, in the ideal scenario, we expect the LLM to judge the correctness of its generated code, assigning 100\% confidence to correct code and 0\% confidence to incorrect code.
        %Given the binary nature of correctness in each generated code sample, in the ideal scenario, we would expect the LLM to accurately judge the correctness of its generated code, assigning 100\% confidence to correct code and 0\% confidence to incorrect code. 
        Formally~\cite{jiang2021knowlanguagemodelsknow}: for a model $M$, given an input $x$ and the set $\mathbf{Y}$ of all correct outputs for this input, the model produces an output $M(x) = y$ and assigns a confidence $p(y | x) \in [0, 1]$ to this output. We define 
        \begin{equation}
            \delta(y | x) = \begin{cases} 1, & y \in \mathbf{Y} \\ 0, & y \notin \mathbf{Y} \end{cases}
        \end{equation}
        where $\delta(y | x)$  denotes whether the model's output is correct. 
        In the ideal scenario, $\forall x, p(y | x) = \delta(y | x)$. A model that satisfies this property is referred to as \textbf{well-calibrated}.

        However, as no LLM is currently perfect, we take a step back from the ideal scenario. For a given set of inputs $\mathbf{X}$, we examine the model's average confidence performance across the entire set. We expect this average to be close to the actual value, rather than requiring each individual $p(y | x)$ to be equal to each $\delta(y | x)$. Specifically, we examine
        \begin{equation}
            p_{\mathbf{X}} = \frac{1}{|\mathbf{X}|} \sum_{x_{i} \in \mathbf{X}} p(y_{i} | x_{i}),
            \delta_{\mathbf{X}} = \frac{1}{|\mathbf{X}|} \sum_{x_{i} \in \mathbf{X}} \delta(y_{i} | x_{i})
        \end{equation}
        and hope that the values of $p_{\mathbf{X}}$ and $\delta_{\mathbf{X}}$ are as close as possible. We consider the confidence generated by an LLM to be \textbf{reliable} if the values of $p_{\mathbf{X}}$ and $\delta_{\mathbf{X}}$ remain consistently close across different tasks (i.e., different $\mathbf{X}$).

        Therefore, this study conducts an empirical investigation into the reliability of confidence in current mainstream LLMs on typical code reasoning task benchmarks, and investigates the effect of different processing methods on improving the confidence reliability of LLMs. %The objective is to determine whether there exist technical approaches that would enable \textbf{confidence} to be adopted in practical development, thereby providing valuable technical guidance for software engineering.

%% file: sections/3methodology.tex
\section{Methodology}

    In this section, we present the methodology of this study. Figure \ref{fig:methodology} illustrates an overview of the methodology, followed by a detailed discussion of each step.

    \subsection{Overview}

        \begin{figure*}
            \centering
            \includegraphics[width=\textwidth]{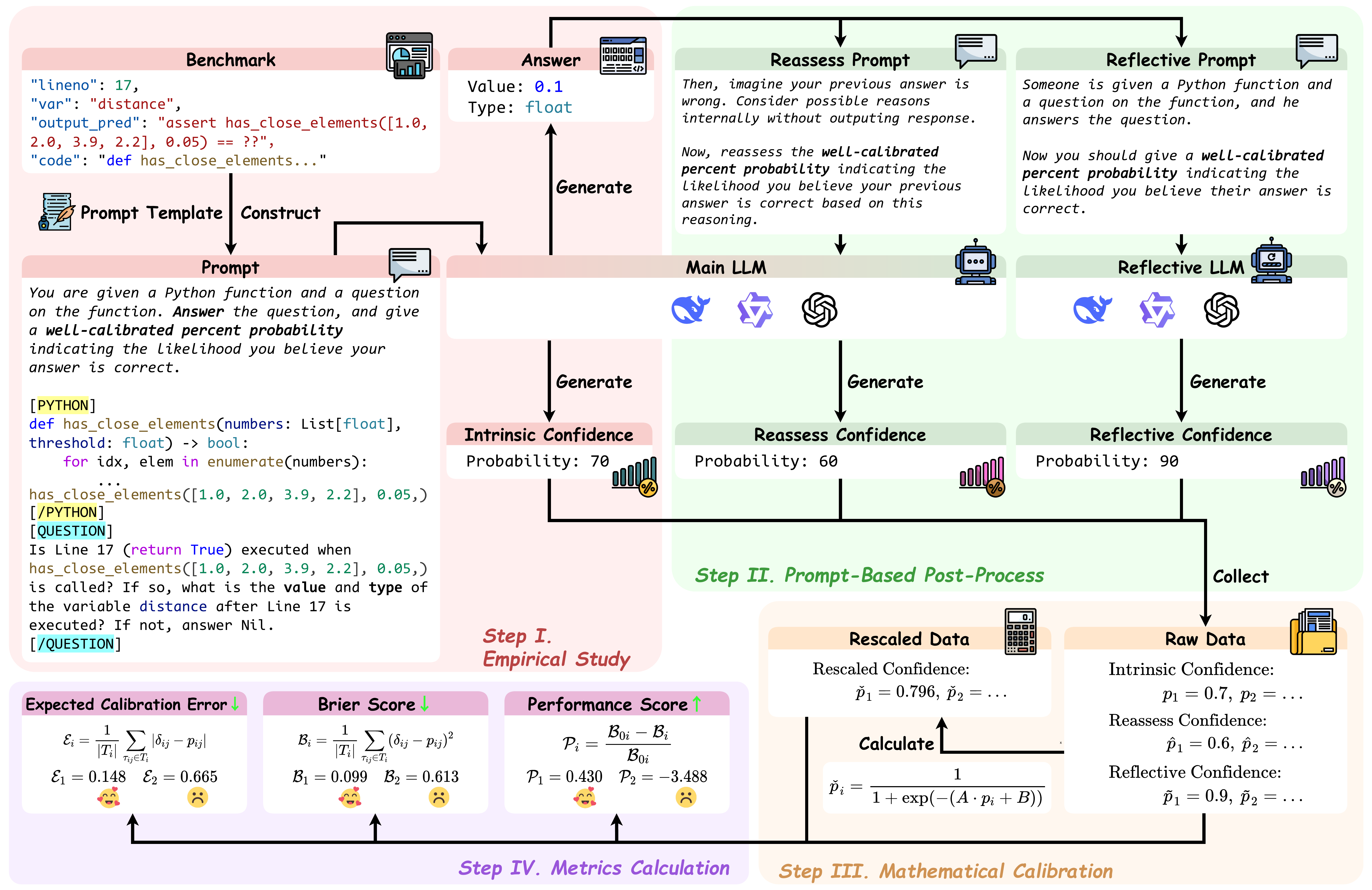}
            \caption{Overview of the entire research process.}
            \vspace{-0.3cm}
            \label{fig:methodology}
        \end{figure*}

        As illustrated in Figure~\ref{fig:methodology}, this process is divided into four steps, each playing a distinct role throughout the workflow:

        \begin{itemize}
            \item \textbf{Step 1. Empirical Study}. Prompt LLMs to generate both answers to test cases and their corresponding confidence scores, which not only constitutes the fundamental empirical study but also provides baseline results for subsequent research.
            \item \textbf{Step 2. Prompting Strategy Adjustment}. Employ different prompt strategies to prompt LLMs to generate confidence scores for their answers once again.
            \item \textbf{Step 3. Mathematical Calibration}. Apply mathematical methods to process the confidence scores generated by LLMs.
            \item \textbf{Step 4. Metrics Calculation}. Compute various metrics to evaluate the reliability of the different types of confidence scores obtained in the previous stages.
        \end{itemize}

    \subsection{Empirical Study}

        Our research begins with an empirical study on the reliability of confidence estimates produced by mainstream LLMs on code reasoning task benchmarks. To ensure the validity of our study, we select a wide range of code reasoning task benchmarks and design prompt templates to standardise the handling of different test questions\footnote{For specific prompts, see Figure~\ref{fig:methodology}}. These prompts are then provided to the LLMs, which are required to give an answer to each question, as well as a confidence score for their answer.

        Formally, we denote the test set for each subtask as $T_{i}$, where each test point within the test set is denoted as $\tau_{ij} \in T_{i}$. For each test point, we define:
        \begin{equation}
            \delta_{ij} = \begin{cases}
                1, & \text{answer is correct} \\
                0, & \text{answer is incorrect}
            \end{cases}
        \end{equation}
        to indicate the correctness of the LLM's result on this test point.

        To distinguish the confidence generated at this step from those produced in subsequent steps, we refer to the confidence generated here as \textbf{intrinsic} confidence, denoted as $p_{ij}$. By evaluating the correctness of the LLMs' answers and their \textbf{intrinsic} confidence scores, we can assess the reliability of confidence estimates produced by mainstream LLMs on code reasoning tasks without any additional measures, thus completing the empirical study. The results of the empirical study will also serve as baseline data, to be compared with the confidence scores obtained after introducing additional measures in subsequent sections, in order to evaluate the improvement in confidence reliability brought about by these measures.

    \subsection{Prompt Strategy Optimisation}

        In the \textbf{prompt strategy optimisation} step, we employ some prompt strategies to obtain the confidence scores that LLMs assign to their answers on code reasoning tasks. Subsequently, we evaluate these strategies using relevant metrics to determine whether they enhance the reliability of confidence estimates. Specifically, we propose two distinct prompt strategies, namely the \textbf{reassess strategy} and the \textbf{reflective strategy}.

        \noindent \textbf{Reassess Strategy}. After the LLM generates the \textbf{intrinsic} confidence score alongside its answer for each test point, we invoke the same LLM again, requesting it to reassess the confidence in its previously generated answer. In particular, to ensure the effectiveness of this reassessment and to improve the confidence estimate, we adopt a self-doubt strategy, whereby the LLM is prompted to provide a new confidence score under the assumption that its previous answer may not necessarily be correct\footnote{For specific prompts, see Figure~\ref{fig:methodology}}. Some studies have shown that such an approach helps to mitigate the overconfidence problem in LLMs and can implicitly increase the depth of their reasoning~\cite{xiong2024can, yu2024fightfiretrustchatgpt}. The confidence score generated under this strategy is referred to as the \textbf{reassess} confidence, denoted as $\hat{p}_{ij}$.

        \noindent \textbf{Reflective Strategy}. After obtaining the answer from the main LLM, we provide both the question and the answer to a newly invoked reflective LLM, asking it to assign a confidence score to the main LLM's answer\footnote{For specific prompts, see Figure~\ref{fig:methodology}}. Although both LLMs are driven by the same underlying model, there is no shared context between them. Thus, the reflective LLM acts purely as an evaluator rather than as a respondent. This approach helps to avoid the situation where the answering LLM, in an attempt to defend its own answer or to accommodate the user, evaluates confidence from a less than fully objective perspective~\cite{spiess2024calibrationcorrectnesslanguagemodels}. The confidence score generated under this strategy is referred to as the \textbf{reflective} confidence, denoted as $\tilde{p}_{ij}$.

    \subsection{Mathematical Calibration}

        In addition to prompt adjustment, another potential method for improving the reliability of confidence scores is \textbf{mathematical calibration} of the raw confidence data, making it more closely aligned with the actual correctness of the LLM's answers.

        In this study, the data under consideration is characterised by its binary nature (correct or incorrect). For such binary data, a variety of calibration methods are available~\cite{guo2017calibrationmodernneuralnetworks}. However, these methods share a common feature: they aggregate the discrete results for individual test points. Their goal is not to achieve precise fitting at each single test point. Instead, they aim for values that, on average, approach the true results within an appropriately sized group of test points. Among the various calibration methods, \textit{Platt Scaling}~\cite{platt1999probabilistic} is widely used in LLM research. Accordingly, this study also adopts \textit{Platt Scaling} for data calibration.

        Formally, we denote the confidence score obtained after applying \textit{Platt Scaling} as $\check{p}_{ij}$, and is calculated as follows:
        \begin{equation}
            \check{p}_{ij} = \frac{1}{1 + \exp(-(A \cdot p_{ij} + B))}
        \end{equation}
        where parameters $A$ and $B$ are optimised to minimise the negative log-likelihood on the calibration data~\cite{platt1999probabilistic}. To prevent overfitting, we implement 5-fold cross-validation:
        \begin{enumerate}
            \item Partition the dataset into five equal subsets.
            \item For each fold, use four subsets for parameter estimation.
            \item Apply the learned parameters to the held-out subset.
            \item Aggregate results across all folds.
        \end{enumerate}
        This approach ensures robust calibration while maintaining the statistical validity of our results~\cite{spiess2024calibrationcorrectnesslanguagemodels}.

    \subsection{Metrics Calculation}

        After the aforementioned stages, we obtain the \textbf{intrinsic} confidence $p_{ij}$, \textbf{reassess} confidence $\hat{p}_{ij}$, \textbf{reflective} confidence $\tilde{p}_{ij}$, and \textit{Platt-scaled} confidence $\check{p}_{ij}$. For these different types of confidence scores, we employ the same set of evaluation metrics. Based on these metric results, we conduct an empirical study on the reliability of confidence estimates across different models, and further assess the improvements in confidence reliability brought about by various methods. In the following, we use $p_{ij}$ as an example to illustrate the evaluation metrics.

        \subsubsection{Expected Calibration Error}

            \textbf{Expected Calibration Error} (\textbf{ECE})~\cite{naeini2015obtaining} is a typical metric used to reflect the accuracy of the confidence of LLMs. Let the \textbf{ECE} for each subtask be denoted as $\mathcal{E}_{i}$, defined as:
            \begin{equation}
                \mathcal{E}_{i} = \frac{1}{|T_{i}|} \sum_{\tau_{ij} \in T_{i}} |\delta_{ij} - p_{ij}|
            \end{equation}

            It is evident from the definition that the higher the accuracy of the confidence, the smaller the value of $\mathcal{E}_{i}$.

        \subsubsection{Brier Score}

            While \textbf{ECE} is intuitive, many studies have pointed out a limitation of this metric: when a relatively weak LLM cannot accurately predict the correctness of its own answers and instead always outputs a baseline value, a low \textbf{ECE} value may still be obtained~\cite{spiess2024calibrationcorrectnesslanguagemodels}. To address this, we introduce another metric, the \textbf{Brier Score} (\textbf{BS})~\cite{brier1950verification}. Let the \textbf{Brier Score} for each subtask be denoted as $\mathcal{B}_{i}$, defined as:
            \begin{equation}
                \mathcal{B}_{i} = \frac{1}{|T_{i}|} \sum_{\tau_{ij} \in T_{i}} (\delta_{ij} - p_{ij})^{2}
            \end{equation}

            By definition, the higher the accuracy of the confidence, the smaller the value of $\mathcal{B}_{i}$, and only in the ideal perfect case, $\mathcal{B}_{i} = 0$. Thus, this metric effectively addresses the issue caused by outputting a baseline value.

        \subsubsection{Performance Score}

            For a more intuitive comparison, we further define the \textbf{Performance Score} (\textbf{PS}), which reflects the intelligence level of the LLM in making predictions. Specifically, for each subtask, we compute the average confidence $p_{ij}$ across all data points, denoted as $\bar{p_{i}}$, and use this value as the baseline confidence for the model on that subtask. We then use this baseline to compute a \textbf{Brier Score}, denoted as $\mathcal{B}_{0i}$, with the closed form:
            \begin{equation}
                \mathcal{B}_{0i} = \bar{p_{i}} (1 - \bar{p_{i}})
            \end{equation}

            Based on $\mathcal{B}_{0i}$, we define the \textbf{Performance Score} for the subtask as $\mathcal{P}_{i}$:
            \begin{equation}
                \mathcal{P}_{i} = \frac{\mathcal{B}_{0i} - \mathcal{B}_{i}}{\mathcal{B}_{0i}}
            \end{equation}

            According to the definition, \textbf{Performance Score} effectively reflects the intelligence of the LLM compared to guessing based only on the baseline value. Specifically, in the ideal perfect case, $\mathcal{P}_{i} = 1$; when $\mathcal{P}_{i}$ is positive, it indicates that the LLM performs better than guessing based only on the baseline value; when $\mathcal{P}_{i}$ is negative, it indicates that the LLM performs worse than guessing based only on the baseline value.

        In summary, the three representative metrics, \textbf{ECE}, \textbf{Brier Score}, and \textbf{Performance Score}, can effectively characterise the reliability of LLM confidence estimates for each subtask, while excluding misleading results. The evaluation of \textbf{intrinsic} confidence constitutes the empirical study. The assessments of \textbf{reassess} confidence and \textbf{reflective} confidence demonstrate the effectiveness of different prompt strategies. The evaluation of confidence scores after \textit{Platt Scaling} reflects the effectiveness of mathematical calibration.

%% file: sections/4setup.tex
\section{Experimental Setup}

    This section introduces the specific research questions addressed in this study, as well as the benchmarks and LLMs selected on this basis.

    \subsection{Research Questions}

        Based on the aforementioned research background and methodological introduction, we specify the content of this study in the form of the following three research questions:

        \begin{itemize}
            \item \textbf{RQ1}: How reliable is the confidence of large language models on code reasoning tasks?
            \item \textbf{RQ2}: To what extent can \textbf{prompt strategy optimisation} methods improve the reliability of confidence?
            \item \textbf{RQ3}: To what extent can \textbf{mathematical calibration} methods improve the reliability of confidence?
        \end{itemize}

        Centred around these three research questions, we select appropriate benchmarks and LLMs to ensure the significance and validity of the research findings.

    \subsection{Benchmarks and Tasks}

        \begin{table}[h]
            \centering
            \caption{Overall statistics of all benchmarks used in this study.}
            \vspace{-0.2cm}
           \resizebox{\linewidth}{!}{ \begin{tabular}{|c|c|c|}
                \hline
                \textbf{Benchmark} & \textbf{Benchmark Size} & \textbf{Task} \\
                \hline
                \multirow{4}{*}{\textit{REval}} & \multirow{4}{*}{3,152} 
                    & Code Coverage Prediction (\textbf{CCP}) \\
                    & & Programme State Prediction (\textbf{PSP}) \\
                    & & Execution Path Prediction (\textbf{EPP}) \\
                    & & Output Prediction (\textbf{OP}) \\
                \hline
                \multirow{2}{*}{\textit{CRUXEval}} & \multirow{2}{*}{800} 
                    & Input Prediction (\textbf{CRUXEval-I}) \\
                    & & Output Prediction (\textbf{CRUXEval-O}) \\
                \hline
            \end{tabular}}
            \label{tab:datasets}
        \end{table}

        \subsubsection{REval~\cite{chen2025reasoning}}

            \textit{REval} is a benchmark designed to evaluate the code reasoning abilities of LLMs, with a focus on assessing whether LLMs can correctly reason about intermediate behaviours during programme execution. Specifically, \textit{REval} comprises four subtasks:
            \begin{enumerate}
                \item \textbf{Code Coverage Prediction} (\textbf{CCP}): reasoning whether a statement in the code will be executed under a given input.
                \item \textbf{Programme State Prediction} (\textbf{PSP}): reasoning about the type and value of a specified variable in the code.
                \item \textbf{Execution Path Prediction} (\textbf{EPP}): given a breakpoint during code execution, reasoning about the next statement to be executed.
                \item \textbf{Output Prediction} (\textbf{OP}): reasoning about the output of the code under a given input.
            \end{enumerate}

            \textit{REval} is based on 154 function codes, with several test points designed for each function, resulting in a total of 3,152 test points. Its large scale and the richness of function features ensure that this benchmark provides a comprehensive assessment of LLMs' ability to reason about intermediate behaviours during programme execution, as confirmed by related research~\cite{hu2025assessingadvancingbenchmarksevaluating}.

        \subsubsection{CRUXEval}

            \textit{CRUXEval}~\cite{gu2024cruxevalbenchmarkcodereasoning} is also a benchmark for evaluating the code reasoning abilities of LLMs, focusing on whether LLMs can correctly reason about function input-output pairs. Specifically, \textit{CRUXEval} includes two subtasks:
            \begin{enumerate}
                \item \textbf{Input Prediction} (\textbf{CRUXEval-I}): given the output of a function, reasoning about the possible input to the function.
                \item \textbf{Output Prediction} (\textbf{CRUXEval-O}): given the input to a function, reasoning about the output of the function.
            \end{enumerate}

            \textit{CRUXEval} consists of 800 independent Python functions, each with specifically constructed input-output pairs. The construction and evaluation methods of this benchmark are both representative~\cite{xu2025cruxevalxbenchmarkmultilingualcode}, and its effectiveness in evaluating the reasoning abilities of LLMs has been validated in numerous studies~\cite{hu2025assessingadvancingbenchmarksevaluating, zhu2025evolutionaryperspectivesevaluationllmbased, zhao2025learningreasonexternalrewards}. Furthermore, given the representativeness of \textit{CRUXEval} in code reasoning tasks, many LLMs have cited their performance on this benchmark in their technical reports~\cite{yin2025panguultrapushinglimits, lozhkov2024starcoder2stackv2, hui2024qwen25codertechnicalreport, yang2025qwen3technicalreport}. Therefore, we are confident that \textit{CRUXEval} provides an effective and comprehensive evaluation of LLMs' code reasoning abilities.

    \subsection{Models}

        We select a range of LLMs that are currently widely used in software engineering development, including: \textit{DeepSeek-V3}~\cite{deepseekai2025deepseekv3technicalreport}, \textit{DeepSeek-R1}~\cite{deepseekai2025deepseekr1incentivizingreasoningcapability}, \textit{Qwen3} series~\cite{yang2025qwen3technicalreport} (i.e., \textit{Qwen3-1.7B}, \textit{Qwen3-14B}, and \textit{Qwen3-32B}) and \textit{GPT-3.5 Turbo}~\footnote{\url{https://platform.openai.com/docs/models/gpt-3.5-turbo}}.

        The selection of these LLMs is motivated by the following comparative considerations:
        \begin{enumerate}
            \item \textbf{Reasoning vs Non-reasoning}. By comparing the performance of \textit{DeepSeek-V3} and \textit{DeepSeek-R1}, we investigate the impact of reasoning mechanisms in LLMs on the reliability of confidence in code reasoning tasks.
            \item \textbf{Small-scale vs Large-scale}. By comparing LLMs of different sizes within the \textit{Qwen3} series, we examine the effect of training data scale on the reliability of confidence in code reasoning.
            \item \textbf{Open-source vs Closed-source}. By comparing the open-source \textit{DeepSeek} and \textit{Qwen3} series models with the closed-source \textit{GPT-3.5} series models, this study investigates the performance differences in confidence reliability between mainstream open-source and closed-source models on code reasoning tasks in practical development.
        \end{enumerate}
        We believe that this selection covers the main characteristics of current mainstream LLMs, providing better insights for cutting-edge development.
        In the specific experiments, given the precise and rational nature of code reasoning tasks, we set the temperature parameter of all models (where configurable) to 0 to ensure the stability of generated answers.

%% file: sections/5results.tex
\section{Results}

    In this section, we discuss the research findings by addressing three research questions.

    \begin{table*}
        % \scriptsize
        % \setlength{\tabcolsep}{1.1pt}
        % \renewcommand{\arraystretch}{1.1}
        \caption{The reliability performance of LLMs confidence on code reasoning tasks.}
        \vspace{-0.2cm}
        \resizebox{\linewidth}{!}{
        \centering
        \begin{tabular}{|c|c|ccc|ccc|ccc|ccc|ccc|ccc|ccc|}
        \hline
        \multirow{3}{*}{Model} & \multirow{3}{*}{Type} 
        & \multicolumn{12}{c|}{REval} 
        & \multicolumn{6}{c|}{CRUXEval} \\
        \cline{3-20}
        & & \multicolumn{3}{c|}{\textbf{CCP}} & \multicolumn{3}{c|}{\textbf{PSP}} & \multicolumn{3}{c|}{\textbf{EPP}} & \multicolumn{3}{c|}{\textbf{OP}} 
        & \multicolumn{3}{c|}{\textbf{CRUXEval-I}} & \multicolumn{3}{c|}{\textbf{CRUXEval-O}} \\
        \cline{3-20}
        & & ECE $\downarrow$ & BS $\downarrow$ & PS $\uparrow$ & ECE $\downarrow$ & BS $\downarrow$ & PS $\uparrow$ & ECE $\downarrow$ & BS $\downarrow$ & PS $\uparrow$ & ECE $\downarrow$ & BS $\downarrow$ & PS $\uparrow$ & ECE $\downarrow$ & BS $\downarrow$ & PS $\uparrow$ & ECE $\downarrow$ & BS $\downarrow$ & PS $\uparrow$ \\
        \hline

        % DeepSeek-Chat
        \multirow{3}{*}{DeepSeek-Chat}
        & intrinsic  & \databg{0.143} & \databg{0.093} & \databg{-0.795} & \databg{0.240} & \databg{0.182} & \databg{-2.034} & \databg{0.381} & \databg{0.326} & \databg{-4.637} & \databg{0.170} & \databg{0.105} & \databg{-0.500} & \databg{0.301} & \databg{0.201} & \databg{-0.817} & \databg{0.337} & \databg{0.257} & \databg{-1.886} \\
        & reassess   & \databg{0.233} & \databg{0.126} & \databg{0.093} & \databg{0.353} & \databg{0.212} & \databg{-0.133} & \databg{0.519} & \databg{0.334} & \databg{-0.388} & \databg{0.218} & \databg{0.099} & \databg{0.283} & \databg{0.405} & \databg{0.203} & \databg{0.145} & \databg{0.388} & \databg{0.194} & \databg{0.127} \\
        & reflective & \databg{0.171} & \databg{0.095} & \databg{-0.100} & \databg{0.260} & \databg{0.171} & \databg{-0.638} & \databg{0.371} & \databg{0.272} & \databg{-1.414} & \databg{0.171} & \databg{0.110} & \databg{-0.515} & \databg{0.318} & \databg{0.205} & \databg{-0.431} & \databg{0.333} & \databg{0.237} & \databg{-1.019} \\
        \hline

        % DeepSeek-Reasoner
        \multirow{3}{*}{DeepSeek-Reasoner}
        & intrinsic  & \databg{0.066} & \databg{0.032} & \databg{0.086} & \databg{0.148} & \databg{0.099} & \databg{-0.982} & \databg{0.286} & \databg{0.232} & \databg{-2.813} & \databg{0.072} & \databg{0.027} & \databg{0.430} & \databg{0.045} & \databg{0.018} & \databg{0.376} & \databg{0.038} & \databg{0.017} & \databg{0.258} \\
        & reassess   & \databg{0.177} & \databg{0.105} & \databg{0.261} & \databg{0.224} & \databg{0.125} & \databg{0.225} & \databg{0.320} & \databg{0.234} & \databg{-0.570} & \databg{0.147} & \databg{0.053} & \databg{0.435} & \databg{0.076} & \databg{0.025} & \databg{0.633} & \databg{0.072} & \databg{0.017} & \databg{0.752} \\
        & reflective & \databg{0.130} & \databg{0.084} & \databg{0.186} & \databg{0.169} & \databg{0.117} & \databg{-0.212} & \databg{0.298} & \databg{0.252} & \databg{-1.561} & \databg{0.162} & \databg{0.064} & \databg{0.469} & \databg{0.046} & \databg{0.024} & \databg{0.472} & \databg{0.047} & \databg{0.014} & \databg{0.702} \\
        \hline

        % Qwen3-1.7B
        \multirow{3}{*}{Qwen3-1.7B}
        & intrinsic  & \databg{0.231} & \databg{0.185} & \databg{-1.829} & \databg{0.454} & \databg{0.407} & \databg{-5.338} & \databg{0.663} & \databg{0.577} & \databg{-5.115} & \databg{0.326} & \databg{0.303} & \databg{-7.096} & \databg{0.572} & \databg{0.384} & \databg{-1.023} & \databg{0.573} & \databg{0.418} & \databg{-1.512} \\
        & reassess   & \databg{0.272} & \databg{0.153} & \databg{-0.257} & \databg{0.470} & \databg{0.336} & \databg{-1.328} & \databg{0.666} & \databg{0.519} & \databg{-2.108} & \databg{0.480} & \databg{0.349} & \databg{-1.455} & \databg{0.584} & \databg{0.393} & \databg{-1.020} & \databg{0.574} & \databg{0.418} & \databg{-1.509} \\
        & reflective & \databg{0.275} & \databg{0.154} & \databg{-0.198} & \databg{0.485} & \databg{0.359} & \databg{-1.712} & \databg{0.617} & \databg{0.481} & \databg{-2.329} & \databg{0.501} & \databg{0.383} & \databg{-2.118} & \databg{0.599} & \databg{0.448} & \databg{-1.821} & \databg{0.642} & \databg{0.555} & \databg{-5.024} \\
        \hline

        % Qwen3-14B
        \multirow{3}{*}{Qwen3-14B}
        & intrinsic  & \databg{0.096} & \databg{0.048} & \databg{0.009} & \databg{0.176} & \databg{0.121} & \databg{-1.098} & \databg{0.378} & \databg{0.319} & \databg{-4.179} & \databg{0.101} & \databg{0.048} & \databg{0.107} & \databg{0.453} & \databg{0.388} & \databg{-4.622} & \databg{0.485} & \databg{0.420} & \databg{-5.022} \\
        & reassess   & \databg{0.497} & \databg{0.285} & \databg{-0.140} & \databg{0.548} & \databg{0.350} & \databg{-0.422} & \databg{0.501} & \databg{0.290} & \databg{-0.158} & \databg{0.285} & \databg{0.114} & \databg{0.435} & \databg{0.509} & \databg{0.299} & \databg{-0.199} & \databg{0.473} & \databg{0.271} & \databg{-0.189} \\
        & reflective & \databg{0.209} & \databg{0.081} & \databg{0.447} & \databg{0.275} & \databg{0.139} & \databg{0.158} & \databg{0.384} & \databg{0.307} & \databg{-2.491} & \databg{0.162} & \databg{0.064} & \databg{0.469} & \databg{0.436} & \databg{0.276} & \databg{-0.520} & \databg{0.429} & \databg{0.300} & \databg{-0.954} \\
        \hline

        % Qwen3-32B
        \multirow{3}{*}{Qwen3-32B}
        & intrinsic  & \databg{0.109} & \databg{0.046} & \databg{0.304} & \databg{0.178} & \databg{0.114} & \databg{-0.707} & \databg{0.379} & \databg{0.298} & \databg{-2.442} & \databg{0.105} & \databg{0.046} & \databg{0.282} & \databg{0.384} & \databg{0.315} & \databg{-3.175} & \databg{0.419} & \databg{0.317} & \databg{-1.748} \\
        & reassess   & \databg{0.472} & \databg{0.256} & \databg{-0.027} & \databg{0.617} & \databg{0.416} & \databg{-0.827} & \databg{0.547} & \databg{0.339} & \databg{-0.524} & \databg{0.310} & \databg{0.139} & \databg{0.337} & \databg{0.386} & \databg{0.260} & \databg{-0.356} & \databg{0.380} & \databg{0.260} & \databg{-0.405} \\
        & reflective & \databg{0.167} & \databg{0.074} & \databg{0.386} & \databg{0.207} & \databg{0.115} & \databg{0.106} & \databg{0.372} & \databg{0.290} & \databg{-1.593} & \databg{0.147} & \databg{0.072} & \databg{0.292} & \databg{0.368} & \databg{0.285} & \databg{-0.984} & \databg{0.375} & \databg{0.267} & \databg{-0.671} \\
        \hline

        % GPT-3.5-Turbo
        \multirow{3}{*}{GPT-3.5-Turbo}
        & intrinsic  & \databg{0.338} & \databg{0.317} & \databg{-13.566} & \databg{0.665} & \databg{0.613} & \databg{-10.510} & \databg{0.807} & \databg{0.764} & \databg{-15.122} & \databg{0.335} & \databg{0.277} & \databg{-3.488} & \databg{0.725} & \databg{0.654} & \databg{-7.580} & \databg{0.534} & \databg{0.477} & \databg{-6.837} \\
        & reassess   & \databg{0.448} & \databg{0.255} & \databg{-0.051} & \databg{0.513} & \databg{0.330} & \databg{-0.331} & \databg{0.610} & \databg{0.415} & \databg{-0.817} & \databg{0.430} & \databg{0.234} & \databg{0.039} & \databg{0.574} & \databg{0.364} & \databg{-0.576} & \databg{0.514} & \databg{0.352} & \databg{-0.733} \\
        & reflective & \databg{0.353} & \databg{0.315} & \databg{-1.896} & \databg{0.610} & \databg{0.522} & \databg{-2.860} & \databg{0.767} & \databg{0.702} & \databg{-6.830} & \databg{0.408} & \databg{0.300} & \databg{-0.955} & \databg{0.438} & \databg{0.298} & \databg{-0.195} & \databg{0.497} & \databg{0.427} & \databg{-2.250} \\
        \hline
        \end{tabular}}
        \label{tab:raw}
    \end{table*}

    \subsection{RQ1: How reliable is the confidence of large language models on code reasoning tasks?}

        The \textbf{intrinsic} confidence of LLMs on each subtask can effectively characterise the reliability of their confidence in code reasoning tasks. Table~\ref{tab:raw} presents the detailed results.

        \subsubsection{Overall Reliability Analysis}

            Overall, there are significant differences in the reliability of confidence among current mainstream LLMs on code reasoning tasks. In terms of the ECE metric, the intrinsic confidence ECE of \textit{DeepSeek-Reasoner} on the CCP subtask is only 0.066, which is much lower than that of \textit{DeepSeek-Chat} (0.143), \textit{Qwen3-1.7B} (0.231), and \textit{GPT-3.5-Turbo} (0.338), demonstrating superior confidence calibration. Similarly, \textit{Qwen3-32B} achieves an intrinsic confidence ECE of 0.384 on the CRUXEval-I subtask, outperforming \textit{Qwen3-14B} (0.453) and \textit{Qwen3-1.7B} (0.572), though still not as good as \textit{DeepSeek-Reasoner} (0.045). This indicates that models with explicit reasoning capabilities (such as \textit{DeepSeek-Reasoner}) have a clear advantage in confidence reliability.

            The Brier Score further reveals the mean squared error between confidence and actual correctness. \textit{DeepSeek-Reasoner} achieves an intrinsic confidence BS of 0.032 on CCP, lower than \textit{DeepSeek-Chat} (0.093) and \textit{GPT-3.5-Turbo} (0.317), reflecting more accurate confidence prediction. Notably, \textit{GPT-3.5-Turbo} exhibits the highest BS across most subtasks, reaching as high as 0.764 on the EPP subtask, indicating unreliable confidence.

            The Performance Score reflects the intelligence of the model's confidence prediction. Ideally, PS should be positive and as close to 1 as possible. \textit{DeepSeek-Reasoner} achieves an intrinsic confidence PS of 0.376 on the CRUXEval-I subtask, significantly higher than \textit{DeepSeek-Chat} (-0.817), \textit{Qwen3-1.7B} (-1.023), and \textit{GPT-3.5-Turbo} (-7.580), indicating not only good confidence calibration but also effective discrimination between correct and incorrect answers. In contrast, \textit{GPT-3.5-Turbo} has a PS as low as -13.566 on CCP, suggesting its confidence prediction is even worse than baseline guessing.

        \subsubsection{Comparison Between Different Models}

            From a model comparison perspective, the \textit{DeepSeek-Reasoner} with explicit reasoning capabilities outperforms \textit{DeepSeek-Chat} across all metrics, especially on the two \textit{CRUXEval} subtasks, where the intrinsic ECE is 0.045 and 0.038, BS is 0.018 and 0.017, and PS is 0.376 and 0.258, respectively—the best among all models. This demonstrates that the introduction of reasoning capabilities greatly enhances confidence reliability.

            In the \textit{Qwen3} series of models, a particular phenomenon was observed. In the two subtasks of CRUXEval, confidence reliability slightly improves as model size increases. For instance, \textit{Qwen3-32B} achieves an intrinsic confidence ECE and BS of 0.384 and 0.315, respectively, on CRUXEval-I, both outperforming \textit{Qwen3-1.7B} (0.572 and 0.384) and \textit{Qwen3-14B} (0.453 and 0.388). However, across the subtasks of REval, the intrinsic confidence ECE of \textit{Qwen3-14B} is consistently lower than that of \textit{Qwen3-32B}, although the difference is marginal. This suggests that, for specific tasks, the improvement in confidence reliability brought by increasing model size slows down as the scale grows, and may reach a plateau at a certain size, making further enhancement difficult. This implies that smaller-scale models possess remarkable potential in certain code reasoning tasks.

            The closed-source model \textit{GPT-3.5-Turbo} performs poorly in terms of confidence across most subtasks, with the highest intrinsic ECE and BS, and the lowest PS. For example, on the EPP subtask, the intrinsic confidence ECE is 0.807, BS is 0.764, and PS is -15.122, significantly lagging behind open-source models. This suggests that current mainstream open-source LLMs have already surpassed closed-source models in the reliability of confidence for code reasoning.

        \subsubsection{Comparison Between Different Tasks}

            From the perspective of tasks, among the four \textit{REval} subtasks, confidence reliability on CCP and OP is generally better than on PSP and EPP. For example, \textit{DeepSeek-Reasoner} achieves an intrinsic confidence ECE of 0.066 on CCP, which rises to 0.286 on EPP; BS increases from 0.032 to 0.232, and PS drops from 0.086 to -2.813. This indicates that for more complex or diverse intermediate state reasoning tasks, the reliability of model confidence declines significantly.
Among the two CRUXEval subtasks, input prediction (CRUXEval-I) is generally more challenging than output prediction (CRUXEval-O). For instance, for \textit{DeepSeek-Reasoner}, the intrinsic confidence ECE is 0.301 on CRUXEval-I and 0.337 on CRUXEval-O, with a similar trend in BS (from 0.201 to 0.257). This reflects the higher uncertainty of input prediction tasks, making confidence calibration more difficult.These differences indicate significant task-specific trends, with task complexity and the type of reasoning required having a marked impact on confidence and reliability.

        \subsubsection{Brief Summary}

            In summary, there are considerable differences in the reliability of confidence among current mainstream LLMs on code reasoning tasks. Models with reasoning capabilities (such as \textit{DeepSeek-Reasoner}) perform best across all metrics, and open-source models generally outperform closed-source models. The difficulty of different tasks and model scale has a significant impact on confidence reliability, while the improvement brought by different confidence generation methods is limited. Overall, the confidence of current LLMs has not yet reached an ideal level of reliability, and there remains substantial room for improvement, especially for complex reasoning tasks and lower-performing models.

    \subsection{RQ2: To what extent can prompt strategy optimisation methods improve the reliability of confidence?}

        To investigate the effect of prompt strategy optimisation methods on the reliability of confidence, we conducted a comparative analysis of the \textbf{intrinsic}, \textbf{reassess}, and \textbf{reflective} confidence generation approaches across different models and tasks (see Table~\ref{tab:raw}). The \textbf{reassess} strategy prompts the model to reassess its confidence through self-doubt, while the \textbf{reflective} strategy introduces a reflective agent to independently evaluate the main agent’s answer—both theoretically helping to mitigate overconfidence and subjective bias.

        \subsubsection{Overall Improvement Trends}

            Overall, prompt strategy optimisation methods can provide some improvement in confidence reliability for certain models and tasks, but the effect is neither universal nor substantial. For example, for \textit{DeepSeek-Reasoner} on the CRUXEval-O subtask, compared to intrinsic confidence, the reassess confidence ECE decreases from 0.038 to 0.072, BS remains at 0.017, and PS increases from 0.258 to 0.752, indicating a notable positive effect. The reflective confidence on the same task achieves an ECE of 0.047, BS of 0.014, and PS of 0.702, also outperforming intrinsic confidence. This demonstrates that for certain models and tasks, prompt strategy optimisation can further optimise both the intelligence and calibration of confidence. Further analysis reveals that prompt strategy optimisation is more effective for high-performing, reasoning-capable models. For instance, both \textit{DeepSeek-Reasoner} and \textit{Qwen3-32B} show clear improvements in PS on CRUXEval tasks, whereas for lower-performing models such as \textit{Qwen3-1.7B} and \textit{GPT-3.5-Turbo}, the differences among the three confidence generation methods are minimal, and in some cases, PS even turns negative, indicating that the model’s reasoning ability is a prerequisite for reliable confidence.

            However, for most models and tasks, the reassess and reflective strategies do not yield systematic improvements, and in some cases, performance even declines. For example, for \textit{DeepSeek-Chat} on the OP subtask, the reflective confidence ECE increases from 0.170 (intrinsic) to 0.171, and PS drops from -0.500 to -0.515. For \textit{Qwen3-1.7B} across all subtasks, the differences in ECE, BS, and PS between reassess, reflective, and intrinsic confidence are negligible, and overall performance remains poor. Additionally, for some complex tasks such as EPP and CRUXEval-I, the improvement is limited. And in some cases, prompt strategy optimisation may even exacerbate confidence deviation due to the model’s insufficient task understanding. For example, for \textit{Qwen3-1.7B} on CRUXEval-I, the intrinsic PS is only -1.023, and reflective one is -1.821, with no positive improvement. 

        \subsubsection{Confidence Distribution and Calibration Characteristics}

\begin{figure}
                \centering
                \includegraphics[width=0.45\textwidth]{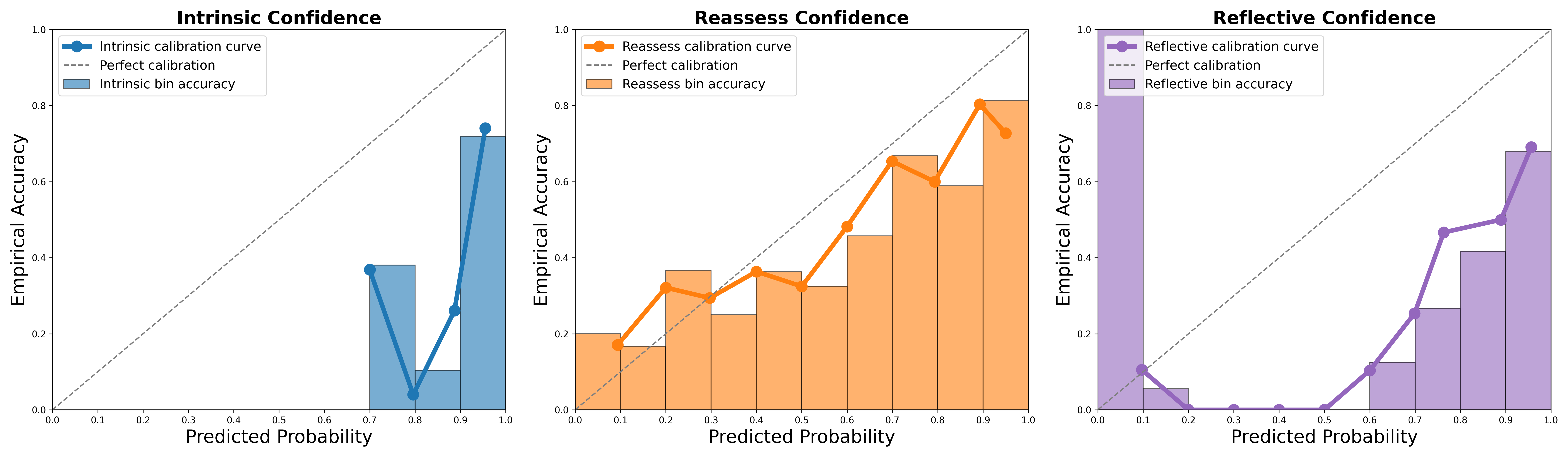}
                \caption{The probability distributions and calibration curves of intrinsic, reassess, and reflective confidence for \textit{GPT-3.5-Turbo} on the OP task.}
                \vspace{-0.3cm}
                \label{fig:rq2}
            \end{figure}

            From the perspective of confidence distribution, the reassess strategy can help alleviate model overconfidence in some cases. For example, for \textit{DeepSeek-Chat} on the CRUXEval-O subtask, the reassess BS decreases from 0.257 (intrinsic) to 0.194, and PS improves from -1.886 to 0.127, indicating that the model’s confidence becomes more aligned with actual correctness after self-doubt. However, for some models, the reflective strategy may introduce new subjectivity, leading to greater confidence fluctuation. A data point reflecting both characteristics is the performance of \textit{GPT-3.5-Turbo} on the OP task, with its probability distribution and calibration curve shown in Figure~\ref{fig:rq2}. It can be seen that, compared to the significantly biased intrinsic confidence, the reassess strategy substantially improves the model’s performance, with its calibration curve closely matching the ideal reference. However, the reflective strategy causes confidence to fluctuate in the opposite direction (towards lower confidence), and its reliability, as shown by the calibration curve, is less promising.

        \subsubsection{Brief Summary}

            In summary, prompt strategy optimisation methods have certain potential for improving the reliability of LLM confidence, particularly for high-performing models and specific tasks. However, their effectiveness is limited by the model’s inherent reasoning ability and task complexity. For low-performing models or complex reasoning tasks, simply relying on prompt strategy optimisation is unlikely to significantly enhance confidence reliability. Therefore, prompt strategy optimisation is better suited as an auxiliary optimisation technique for high-performing LLMs, rather than as a universal solution.

    \subsection{RQ3: To what extent can mathematical calibration methods improve the reliability of confidence?}

        To further enhance the reliability of confidence, we applied the \textit{Platt Scaling} mathematical calibration method to the data of each model, task, and confidence type. Then report the calibrated ECE, BS, and PS metrics in Table~\ref{tab:rescaled}. \textit{Platt Scaling} parameterises the mapping of raw confidence scores to better fit the true accuracy distribution, and in theory, can effectively mitigate systematic biases in model confidence.

        \begin{table*}
            % \scriptsize
            % \setlength{\tabcolsep}{0.6pt}
            % \renewcommand{\arraystretch}{1.1}
            \centering
            \caption{The reliability performance of LLMs confidence on code reasoning tasks after \textit{Platt Scaling}.}
            \vspace{-0.2cm}
           \resizebox{\linewidth}{!}{
            \begin{tabular}{|c|c|ccc|ccc|ccc|ccc|ccc|ccc|ccc|}
            \hline
            \multirow{3}{*}{Model} & \multirow{3}{*}{Type} 
            & \multicolumn{12}{c|}{REval} 
            & \multicolumn{6}{c|}{CRUXEval} \\
            \cline{3-20}
            & & \multicolumn{3}{c|}{\textbf{CCP}} & \multicolumn{3}{c|}{\textbf{PSP}} & \multicolumn{3}{c|}{\textbf{EPP}} & \multicolumn{3}{c|}{\textbf{OP}} 
            & \multicolumn{3}{c|}{\textbf{CRUXEval-I}} & \multicolumn{3}{c|}{\textbf{CRUXEval-O}} \\
            \cline{3-20}
            & & ECE $\downarrow$ & BS $\downarrow$ & PS $\uparrow$ & ECE $\downarrow$ & BS $\downarrow$ & PS $\uparrow$ & ECE $\downarrow$ & BS $\downarrow$ & PS $\uparrow$ & ECE $\downarrow$ & BS $\downarrow$ & PS $\uparrow$ & ECE $\downarrow$ & BS $\downarrow$ & PS $\uparrow$ & ECE $\downarrow$ & BS $\downarrow$ & PS $\uparrow$ \\
            \hline

            % DeepSeek-Chat
            \multirow{3}{*}{DeepSeek-Chat}
            & intrinsic  & \databg{0.181} & \databg{0.091} & \databg{-0.013} & \databg{0.320} & \databg{0.158} & \databg{0.054} & \databg{0.458} & \databg{0.227} & \databg{0.027} & \databg{0.213} & \databg{0.106} & \databg{0.026} & \databg{0.376} & \databg{0.182} & \databg{0.088} & \databg{0.418} & \databg{0.202} & \databg{0.094} \\
            & reassess   & \databg{0.183} & \databg{0.092} & \databg{-0.009} & \databg{0.338} & \databg{0.170} & \databg{-0.006} & \databg{0.461} & \databg{0.233} & \databg{0.008} & \databg{0.205} & \databg{0.101} & \databg{0.091} & \databg{0.381} & \databg{0.189} & \databg{0.061} & \databg{0.404} & \databg{0.198} & \databg{0.117} \\
            & reflective & \databg{0.180} & \databg{0.091} & \databg{-0.010} & \databg{0.322} & \databg{0.162} & \databg{0.037} & \databg{0.424} & \databg{0.211} & \databg{0.099} & \databg{0.217} & \databg{0.108} & \databg{0.034} & \databg{0.387} & \databg{0.192} & \databg{0.046} & \databg{0.408} & \databg{0.197} & \databg{0.117} \\
            \hline

            % DeepSeek-Reasoner
            \multirow{3}{*}{DeepSeek-Reasoner}
            & intrinsic  & \databg{0.063} & \databg{0.032} & \databg{-0.018} & \databg{0.193} & \databg{0.097} & \databg{0.001} & \databg{0.376} & \databg{0.187} & \databg{0.059} & \databg{0.054} & \databg{0.027} & \databg{-0.000} & \databg{0.039} & \databg{0.018} & \databg{0.122} & \databg{0.041} & \databg{0.019} & \databg{0.105} \\
            & reassess   & \databg{0.062} & \databg{0.031} & \databg{0.026} & \databg{0.182} & \databg{0.092} & \databg{0.066} & \databg{0.389} & \databg{0.196} & \databg{0.019} & \databg{0.061} & \databg{0.031} & \databg{0.004} & \databg{0.036} & \databg{0.016} & \databg{0.177} & \databg{0.033} & \databg{0.015} & \databg{0.267} \\
            & reflective & \databg{0.062} & \databg{0.031} & \databg{0.019} & \databg{0.190} & \databg{0.096} & \databg{0.024} & \databg{0.393} & \databg{0.198} & \databg{0.008} & \databg{0.060} & \databg{0.030} & \databg{0.017} & \databg{0.034} & \databg{0.016} & \databg{0.222} & \databg{0.033} & \databg{0.014} & \databg{0.279} \\
            \hline

            % Qwen3-1.7B
            \multirow{3}{*}{Qwen3-1.7B}
            & intrinsic  & \databg{0.291} & \databg{0.148} & \databg{-0.026} & \databg{0.474} & \databg{0.236} & \databg{0.029} & \databg{0.385} & \databg{0.187} & \databg{0.159} & \databg{0.438} & \databg{0.224} & \databg{-0.034} & \databg{0.460} & \databg{0.230} & \databg{0.088} & \databg{0.412} & \databg{0.199} & \databg{0.088} \\
            & reassess   & \databg{0.291} & \databg{0.148} & \databg{-0.029} & \databg{0.496} & \databg{0.249} & \databg{-0.002} & \databg{0.329} & \databg{0.162} & \databg{0.269} & \databg{0.472} & \databg{0.231} & \databg{0.077} & \databg{0.441} & \databg{0.221} & \databg{0.001} & \databg{0.410} & \databg{0.198} & \databg{0.089} \\
            & reflective & \databg{0.291} & \databg{0.148} & \databg{-0.030} & \databg{0.484} & \databg{0.242} & \databg{0.026} & \databg{0.441} & \databg{0.221} & \databg{0.006} & \databg{0.498} & \databg{0.249} & \databg{0.003} & \databg{0.442} & \databg{0.221} & \databg{0.002} & \databg{0.435} & \databg{0.218} & \databg{-0.003} \\
            \hline

            % Qwen3-14B
            \multirow{3}{*}{Qwen3-14B}
            & intrinsic  & \databg{0.095} & \databg{0.048} & \databg{-0.028} & \databg{0.232} & \databg{0.116} & \databg{0.004} & \databg{0.444} & \databg{0.217} & \databg{0.074} & \databg{0.105} & \databg{0.053} & \databg{-0.026} & \databg{0.492} & \databg{0.245} & \databg{0.014} & \databg{0.495} & \databg{0.246} & \databg{0.017} \\
            & reassess   & \databg{0.096} & \databg{0.048} & \databg{-0.007} & \databg{0.238} & \databg{0.120} & \databg{-0.012} & \databg{0.470} & \databg{0.235} & \databg{0.001} & \databg{0.107} & \databg{0.054} & \databg{0.010} & \databg{0.498} & \databg{0.250} & \databg{-0.005} & \databg{0.490} & \databg{0.245} & \databg{0.022} \\
            & reflective & \databg{0.096} & \databg{0.048} & \databg{-0.019} & \databg{0.234} & \databg{0.118} & \databg{0.001} & \databg{0.459} & \databg{0.229} & \databg{0.026} & \databg{0.108} & \databg{0.056} & \databg{-0.015} & \databg{0.468} & \databg{0.230} & \databg{0.075} & \databg{0.425} & \databg{0.200} & \databg{0.201} \\
            \hline

            % Qwen3-32B
            \multirow{3}{*}{Qwen3-32B}
            & intrinsic  & \databg{0.089} & \databg{0.045} & \databg{-0.022} & \databg{0.222} & \databg{0.112} & \databg{-0.003} & \databg{0.431} & \databg{0.212} & \databg{0.100} & \databg{0.097} & \databg{0.049} & \databg{-0.014} & \databg{0.458} & \databg{0.224} & \databg{0.057} & \databg{0.459} & \databg{0.222} & \databg{0.099} \\
            & reassess   & \databg{0.091} & \databg{0.046} & \databg{-0.016} & \databg{0.226} & \databg{0.114} & \databg{-0.014} & \databg{0.470} & \databg{0.235} & \databg{0.002} & \databg{0.106} & \databg{0.053} & \databg{0.018} & \databg{0.452} & \databg{0.226} & \databg{0.052} & \databg{0.438} & \databg{0.216} & \databg{0.125} \\
            & reflective & \databg{0.088} & \databg{0.045} & \databg{-0.012} & \databg{0.203} & \databg{0.103} & \databg{0.029} & \databg{0.450} & \databg{0.225} & \databg{0.041} & \databg{0.104} & \databg{0.053} & \databg{-0.015} & \databg{0.451} & \databg{0.225} & \databg{0.056} & \databg{0.423} & \databg{0.204} & \databg{0.172} \\
            \hline

            % GPT-3.5-Turbo
            \multirow{3}{*}{GPT-3.5-Turbo}
            & intrinsic  & \databg{0.445} & \databg{0.224} & \databg{-0.009} & \databg{0.431} & \databg{0.216} & \databg{-0.005} & \databg{0.267} & \databg{0.136} & \databg{-0.037} & \databg{0.435} & \databg{0.215} & \databg{0.025} & \databg{0.351} & \databg{0.175} & \databg{0.004} & \databg{0.494} & \databg{0.246} & \databg{0.006} \\
            & reassess   & \databg{0.447} & \databg{0.228} & \databg{-0.023} & \databg{0.431} & \databg{0.216} & \databg{-0.007} & \databg{0.266} & \databg{0.136} & \databg{-0.038} & \databg{0.468} & \databg{0.236} & \databg{0.054} & \databg{0.350} & \databg{0.175} & \databg{0.003} & \databg{0.495} & \databg{0.248} & \databg{-0.001} \\
            & reflective & \databg{0.444} & \databg{0.224} & \databg{-0.010} & \databg{0.427} & \databg{0.215} & \databg{-0.003} & \databg{0.266} & \databg{0.136} & \databg{-0.033} & \databg{0.422} & \databg{0.201} & \databg{0.195} & \databg{0.329} & \databg{0.162} & \databg{0.088} & \databg{0.482} & \databg{0.240} & \databg{0.029} \\
            \hline
            \end{tabular}}
            
            \label{tab:rescaled}
        \end{table*}

        \subsubsection{Overall Improvement from Calibration}

            Overall, \textit{Platt Scaling} significantly improves the reliability of confidence across all models and tasks. For example, for \textit{DeepSeek-Chat} on the CCP subtask, the intrinsic confidence ECE decreases from 0.143 to 0.181 after calibration, BS decreases from 0.093 to 0.091, and PS increases from -0.795 to -0.013, almost eliminating negative intelligence. On the CRUXEval-I subtask, PS increases from -0.817 to 0.088, achieving a shift from negative to positive, indicating that the calibrated model confidence now outperforms baseline guessing.

            For high-performing models, mathematical calibration further consolidates their confidence advantage. For instance, for \textit{Qwen3-32B} on the CRUXEval-O subtask, the intrinsic confidence PS increases from -1.748 to 0.099, and BS decreases from 0.317 to 0.222, resulting in a more reasonable confidence distribution. For \textit{DeepSeek-Reasoner} on EPP, PS adjusts from -2.813 to 0.059, and BS decreases from 0.232 to 0.187, making confidence prediction more consistent with reality.

            For lower-performing models, Platt Scaling also brings significant calibration effects. For example, for \textit{Qwen3-1.7B} on the CCP subtask, the intrinsic confidence PS increases from -1.829 to -0.026, and BS decreases from 0.185 to 0.148. Although not fully positive, the systematic bias in confidence is greatly alleviated. For \textit{GPT-3.5-Turbo} on CCP, PS increases from -13.566 to -0.009, and BS decreases from 0.317 to 0.224, indicating that even when the model's confidence is highly unreliable, mathematical calibration can effectively correct its distribution.

        \subsubsection{Calibration Consistency Across Tasks}

            Mathematical calibration demonstrates highly consistent improvement across different tasks. Whether for REval or CRUXEval subtasks, the calibrated ECE and BS of each model decrease significantly, and PS approaches zero or turns positive. This consistency indicates that mathematical calibration methods possess good generalisability and robustness.

        \subsubsection{Synergy Between Prompt Strategy Optimisation and Mathematical Calibration}

            It is noteworthy that after \textit{Platt Scaling} calibration, the differences between intrinsic, reassess and reflective confidence further diminish, with all three converging in terms of ECE, BS, and PS. For example, for \textit{Qwen3-1.7B} on OP, the PS values after \textit{Platt Scaling} for the three confidence types are -0.034, 0.077, and 0.003, all significantly higher than before calibration (-7.096, -1.455, and -2.118) and with minimal differences. This demonstrates that mathematical calibration can effectively eliminate distributional differences arising from different confidence generation methods, making confidence prediction more stable and reliable.

\begin{figure}
                \centering
                \includegraphics[width=0.45\textwidth]{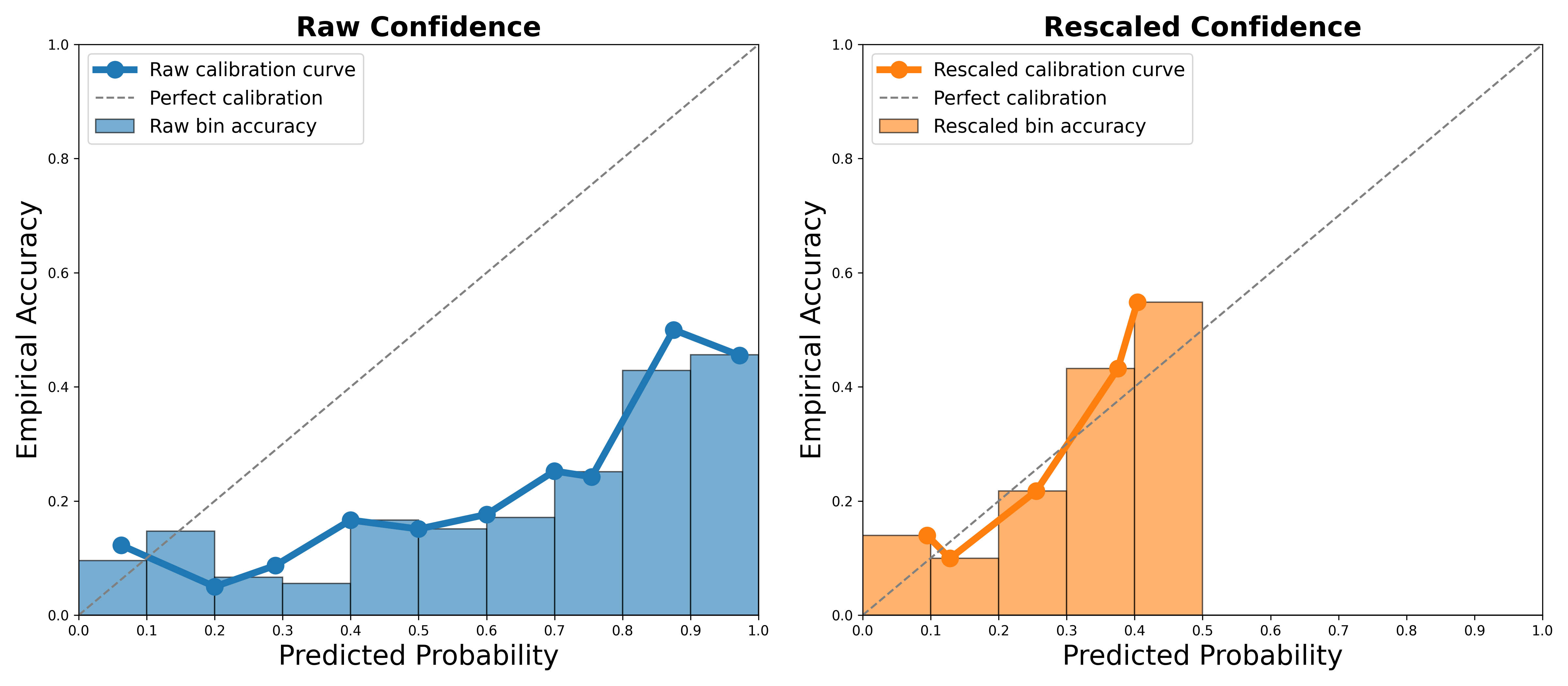}
                \caption{The probability distribution and calibration curve of reflective confidence for \textit{GPT-3.5-Turbo} on the CRUXEval-I task before and after \textit{Platt Scaling}.}
               \vspace{-0.3cm}
                \label{fig:rq3}
            \end{figure}

            Another notable example is \textit{GPT-3.5-Turbo} on the CRUXEval-I task, where the performance of reflective confidence after \textit{Platt Scaling} calibration is illustrated by its probability distribution and calibration curve in Figure~\ref{fig:rq3}. It can be observed that the calibrated probability distribution and curve closely match the ideal curve, in stark contrast to the substantial deviation before calibration.

        \subsubsection{Brief Summary}

            Overall, mathematical calibration methods such as \textit{Platt Scaling} can systematically and significantly improve the reliability of confidence across various models, tasks, and confidence generation methods. Not only do they effectively reduce the bias and mean squared error between confidence and actual accuracy, but they also greatly enhance the intelligence and practical value of model confidence prediction. In comparison, the improvement from prompt strategy optimisation is limited and dependent on the model's inherent capabilities. Therefore, mathematical calibration should be considered the preferred technical approach for improving the reliability of LLM confidence, and can be used in conjunction with prompt strategy optimisation to further optimise model trustworthiness in practical development.

%% file: sections/6discussion.tex
\section{Discussion}

    \subsection{Implications for Researchers}

        Our results reveal that the performance of different LLMs varies significantly, and such differences will directly influence developers’ preferences when selecting LLMs for practical applications. Thus, this study provides LLM developers with a feasible perspective for evaluating the code-related capabilities of their products, namely, by assessing the reliability of their confidence on code reasoning tasks.

        Moreover, we observe that the same LLM may exhibit considerable variation in performance across different code reasoning tasks. On the one hand, this suggests that LLM developers and related researchers should further explore the correlation between task types and model performance. On the other hand, it also provides a practical and targeted direction for improving the code capabilities of LLMs, that is, by prioritising enhancements in tasks where performance is currently suboptimal.

    \subsection{Implications for Software Developers}

        Code reasoning tasks are integral to practical software development processes. With the widespread adoption of LLM-assisted programming, leveraging LLMs to address code reasoning tasks is gradually becoming a paradigm in software development. However, determining the correctness of their reasoning outcomes remains a challenging task~\cite{jesse2023largelanguagemodelssimple}, and uncertainty in this regard may hinder the further advancement of LLM-assisted software development, thereby limiting further improvements in developer productivity. In this study, we propose \textbf{confidence} as a superior method for assessing the reliability of LLMs on code reasoning tasks, and conduct an empirical investigation based on this approach, clearly revealing the capabilities and reliability of current mainstream LLMs in code reasoning tasks. This provides support for software developers regarding the extent to which LLMs can be effectively utilised to assist with code reasoning tasks.
        
        Furthermore, this study explores the effectiveness of two approaches, \textbf{prompt strategy optimisation} and \textbf{mathematical calibration}, in enhancing the reliability of LLM confidence on code reasoning tasks. Some methods, such as the hybrid approach combining the reflective strategy and \textit{Platt Scaling}, significantly improve the reliability of confidence, thereby offering developers effective means to obtain trustworthy confidence estimates. The findings of this study lay the groundwork for the potential effective application of confidence in LLM-assisted code reasoning tasks in the future.

    \subsection{Limitations of Mathematical Calibration Methods}

        Although the section 5 systematically demonstrates the significant effect of mathematical calibration methods in improving the reliability of LLM confidence, there are also notable limitations when considering practical application scenarios. Experimental data show that after \textit{Platt Scaling} calibration, the range of confidence distributions for each model across multiple tasks narrows significantly, with confidence values tending to cluster within certain intervals. While this statistically improves overall calibration and mean squared error, it introduces a new issue: the discriminative power of model confidence decreases, making it more difficult for developers to make nuanced and effective risk judgements based on confidence levels.

        Specifically, a reduced range in confidence means the model’s ability to distinguish between different samples is weakened, with many predictions that were originally high or low confidence being flattened towards the middle after calibration. In practical software engineering, this has two adverse effects. First, developers may find it difficult to intuitively judge which outputs are trustworthy and which require further verification, thus diminishing the utility of confidence as a decision-support tool. Second, in high-risk or critical scenarios, developers need to clearly distinguish between high-confidence outputs and low-confidence outputs, but the convergence of calibrated confidence scores may increase the risk associated with critical decision-making, thereby reducing the practical value of the model.

        Moreover, mathematical calibration methods are essentially global, post-hoc parameter mappings, optimising for overall distributional fit rather than fine-grained discrimination at the individual prediction level. This means that in cases of highly imbalanced sample distributions or tasks with significant heterogeneity, calibration methods may not meet the needs of all subtasks or sample types, and may even obscure the model’s weaknesses in certain specific scenarios. Furthermore, mathematical calibration methods typically rely on sufficient and representative calibration datasets; if there is a distribution shift between the calibration set and the actual application scenario, the calibration effect may be greatly diminished or even introduce new systematic biases.

        Therefore, although mathematical calibration provides an effective technical pathway for improving the reliability of LLM confidence, its limitations in practical software engineering applications should not be overlooked. Future research should further explore how to maintain the discriminative power and interpretability of confidence distributions while improving overall reliability, or combine local calibration and adaptive interval partitioning methods to enhance the practical value and controllability of model confidence in real-world development decisions.

%% file: sections/7threats.tex
\section{Threats to Validity}

    \subsection{Differences Between Benchmarks and Real-World Data}

        Although the datasets in this study cover multiple typical code reasoning scenarios and aim to reflect key challenges faced by LLMs in software engineering, there are inevitable differences between benchmarks and real-world environments. Benchmarks are standardised and structured for reproducibility and comparability, often at the expense of complexity and diversity. In practice, code reasoning tasks can involve more complex logic and ambiguous requirements, which benchmarks may not fully capture. Moreover, benchmarks usually focus on a single reasoning objective, while real-world tasks require balancing multiple objectives. Thus, benchmark results should be cautiously interpreted when applied to real-world scenarios, and the external validity of conclusions is limited~\cite{cao2025buildbenchmarkrevisiting274, hu2025assessingadvancingbenchmarksevaluating}.

    \subsection{Model Selection}

        This study examines mainstream general-purpose LLMs (\textit{DeepSeek}, \textit{Qwen3} series, and \textit{GPT-3.5-Turbo}) to align with industry practices and provide a practical reference. However, this excludes rapidly developing code-specialised LLMs (e.g., \textit{CodeLlama}~\cite{rozière2024codellamaopenfoundation}, \textit{StarCoder}~\cite{lozhkov2024starcoder2stackv2}, and \textit{Magicoder}~\cite{wei2024magicoderempoweringcodegeneration}). These models, pre-trained on code corpora and optimised for code tasks, may show different confidence and reasoning abilities. Differences in training, fine-tuning, and architecture could affect results. Future research should include code-specialised LLMs to better reveal confidence reliability across model types and provide more targeted guidance.

    \subsection{Prompt Design}

        To ensure reproducibility and comparability, this study used a unified and standardised prompt design with minimal flexibility. However, research shows that even small changes in prompt formulation can significantly impact LLM outputs~\cite{luo2014anempiricalanalysis}. Variations in style, context, or order can affect both model accuracy and confidence calibration. Given the vast prompt space and unclear mechanisms, results may vary under different prompt settings. Future work should further investigate the link between prompt engineering and confidence reliability and develop more robust prompt paradigms to enhance generalisability.

    \subsection{Temperature Parameter}

        In this study, to ensure the stability of results, we set the temperature parameter of all LLMs to 0. This aligns with the general understanding of the definition of temperature, namely that a higher temperature may not be suitable for code reasoning tasks, which emphasise logic and precision. However, in certain studies within the field of software engineering~\cite{lee2025generatingdiverse, sun2025sourcecodesummarization, coignion2024aperformancestudyofllmgenerated}, adjusting the temperature parameter has been shown to improve LLM performance on specific tasks. Therefore, the relationship between the temperature parameter and model performance on code reasoning tasks warrants further investigation.

%% file: sections/8related.tex
\section{Related Work}

    \subsection{Confidence Calibration}

        Research on confidence has long been adopted in the modelling field~\cite{brier1950verification}. In the past decade, with the rapid development of LLMs, confidence has also been introduced into LLM research. Initially, this was applied to small neural models, and the effectiveness of improving confidence reliability through scaling~\cite{guo2017calibrationmodernneuralnetworks, srivastava2023beyond} or pre-training~\cite{hendrycks2019pretraining} was observed in these early studies. However, due to the performance limitations of LLMs at that time, such research was still far from practical application environments.
    
        As LLM performance has improved and its applicability to broader task scenarios has increased, confidence research has been applied in a wider range of fields. The work of Jiang \textit{et al.} ~\cite{jiang2021knowlanguagemodelsknow} introduced confidence in the study of LLMs for natural language question answering. Desai \textit{et al.}~\cite{desai2020calibrationpretrainedtransformers} examined the confidence performance of LLMs in natural language reasoning and paraphrasing tasks. Similar studies include the work of Kadavath \textit{et al.}~\cite{kadavath2022languagemodelsmostlyknow}, Xiong \textit{et al.}~\cite{xiong2024can} and Key \textit{et al.}~\cite{key2023trustworthyneuralprogramsynthesis}. In more specialised tasks, Minderer \textit{et al.}~\cite{minderer2021revisitingcalibrationmodernneural} investigated model confidence calibration in computer vision (CV) tasks. Li \textit{et al.}~\cite{li2020Ooerationalcalibrationdebugging} studied operational confidence calibration in computer vision models. Huang \textit{et al.}~\cite{huang2025lookbeforeyouleap} examined the confidence of language models in NLP and function synthesis tasks. Recent research has also explored confidence calibration issues in software domains such as root cause analysis~\cite{zhang2023pacelmpromptingaugmentationcalibrated}. As indicated above, confidence has become a favoured and widely adopted metric in LLM research for natural language tasks.

        Thus far, the beneficial role of well-calibrated confidence in enabling LLMs to judge the correctness of their own answers has been demonstrated~\cite{kadavath2022languagemodelsmostlyknow}. Key \textit{et al.}~\cite{key2023trustworthyneuralprogramsynthesis} have proposed methods for determining whether LLMs can fully solve a problem based on confidence scores. Liu \textit{et al.}~\cite{liu2024enhancinglanguagemodelfactuality} have also explored training strategies for language models based on confidence-related metrics.

    \subsection{Large Language Models in Software Engineering}

        Although the use of LLMs for code-related tasks has been extensively studied in the software engineering field~\cite{zhang2023surveylearningbasedautomatedprogram}, and numerous related benchmarks have been proposed~\cite{cao2025buildbenchmarkrevisiting274, hu2025assessingadvancingbenchmarksevaluating}, most early research focused on code generation or repair, with little attention paid to code reasoning tasks. A representative work is that of Spiess \textit{et al.}\cite{spiess2024calibrationcorrectnesslanguagemodels}, who studied the reliability of LLM confidence in code generation tasks.
    Recently, the software engineering community has begun to focus on code reasoning tasks. Not only have relevant studies emerged~\cite{lamalfa2024codesimulationchallengeslarge}, but new benchmarks have also been proposed~\cite{gu2024cruxevalbenchmarkcodereasoning, xu2025cruxevalxbenchmarkmultilingualcode, chen2025reasoning, liu2025codemindevaluatinglargelanguage}. Nevertheless, research in this area remains relatively limited compared to other aspects.

%% file: sections/9conclusion.tex
\section{Conclusion and Future Work}

    This paper investigates the confidence reliability of LLMs in code reasoning tasks, grounded in practical software engineering scenarios. Our results show that the confidence reliability in mainstream LLMs depends heavily on the model’s reasoning ability, scale, and task type. While larger and more advanced models perform well on simple tasks, their confidence reliability drops for complex tasks or in low-performance scenarios, falling short of practical engineering needs. We further evaluate prompt strategy optimisation and mathematical calibration methods for improving confidence reliability. Their combined use can significantly improve calibration for certain models and tasks, but faces limitations: mathematical calibration often requires targeted data and parameter tuning, limiting generalisability, and may reduce the discriminative power of the confidence distribution, impacting developers’ risk assessment. Enhancing overall reliability while maintaining interpretability and practical utility of confidence scores thus remains a key challenge.

    %To the best of our knowledge, this is the first systematic study of the reliability of LLM confidence on code reasoning tasks. We believe that this research not only provides theoretical foundations and technical references for software engineering practice, but also lays a solid groundwork for further exploration in related fields. 
    Future work may focus on the mechanisms of confidence generation in LLMs for code reasoning tasks, dynamic calibration strategies, and the deep integration of confidence with active learning and risk control techniques. These efforts aim to continually improve the trustworthiness and practicality of LLMs in real-world development, and to fully unlock the application potential of confidence in software engineering.